\documentclass[aps,prl,notitlepage,twocolumn]{revtex4-1}

\usepackage{amsmath}
\usepackage{amssymb}
\usepackage{graphicx}

\begin{document}

\title{Comment on ``Classical Simulations Including Electron Correlations for Sequential Double Ionization''}


\maketitle

In a recent Letter~\cite{zhou12}, the authors investigate trajectories of a two-electron atomic system which models argon submitted to a strong, close to circularly polarized laser pulse. In the field-free case, the following classical Hamiltonian is considered:
\begin{equation}
\label{ham1}
H=\frac{1}{\vert {\bf r}_1-{\bf r}_2\vert }+\sum_{i=1,2} \left[\frac{p_i^2}{2}-\frac{2}{r_i}+V_H(r_i,p_i)\right],
\end{equation}
where ${\bf r}_i \in {\mathbb R}^3$ is the coordinate vector of the $i$-th electron in a frame centered at the nucleus, ${\bf p}_i$ are the canonically conjugate momenta, and $r_i=\vert {\bf r}_i\vert$ and $p_i=\vert {\bf p}_i\vert$. Here the Heisenberg core potential $V_H$ is given by
\begin{equation}
\label{vh}
V_H(r,p)=\frac{\xi^2}{4\alpha r^2}{\rm e}^{\alpha\left( 1-\frac{r^4p^4}{\xi^4}\right)},
\end{equation}
and is supposed to incorporate the Heisenberg uncertainty principle in a classical way~\cite{kirs80}. 
Using Hamiltonian~(\ref{ham1}), the authors of Ref.~\cite{zhou12} analyze the double ionization of argon in order to interpret the experimentally measured ionization times of Ref.~\cite{pfei11}, and in particular the anomalous ionization time of the second electron.  
In a very striking way, Hamiltonian~(\ref{ham1}) is able to capture a shorter than expected ionization time of the second electron in excellent agreement with the experiment. Below, we show that this excellent agreement is only the result of tweaking the parameter $\alpha$ in the potential~(\ref{vh}), and that this result breaks down as $\alpha$ is varied, in contrast with what the authors claim~\cite{zhou12} ``the results of our calculations do not depend on this parameter.'' Their result is therefore not structurally stable and does not hold as $\alpha$ is made arbitrarily large, which is the regime where the potential has been designed (for the constraint $rp\gtrsim \xi$ to be satisfied)~\cite{kirs80}. Commonly used values for $\alpha$ are 4 or 5, which is a compromise between the constraint and the speed of the integration given the stiffness of the problem. In Ref.~\cite{zhou12}, the choice of one of the parameters $(\xi,\alpha)$ relies on the second ionization potential of the atom ($E_2=-1.01$ for argon). More precisely, the minimum of the one electron Hamiltonian (kinetic energy plus Coulomb interaction with the nucleus plus the Heisenberg core potential) has to be equal to $E_2$. As a result, $\xi$ is chosen equal to
 \begin{equation}
 \label{xi}
 \xi=\sqrt{\frac{-2}{E_2\left(1+\frac{1}{2\alpha}\right)}}.
 \end{equation}
The parameter $\alpha$ is chosen arbitrarily, provided that the ground state of the one electron Hamiltonian occurs at a configuration where $rp=\xi$, which requires that $\alpha\gtrsim 1.256$ (the solution of ${\rm e}^x-1-2x=0$). This translates into a minimum value for $\xi$, approximately equal to $1.190$. For $\alpha=2$, the one electron energy minimum occurs at $\xi \approx 1.259$ and not at $\xi=1.225$ as stated in Ref.~\cite{zhou12}. However choosing a value of $\xi$ given by Eq.~(\ref{xi}) results in an empty ground state energy surface at $E_g=-1.59~{\rm a.u.}$ for the two electron system, and thus a different value of $\xi$ has to be considered. In the following, we consider the value chosen in Ref.~\cite{zhou12}, i.e., $\xi=1.225$.   

\begin{figure}
\includegraphics[width = 0.9 \linewidth]{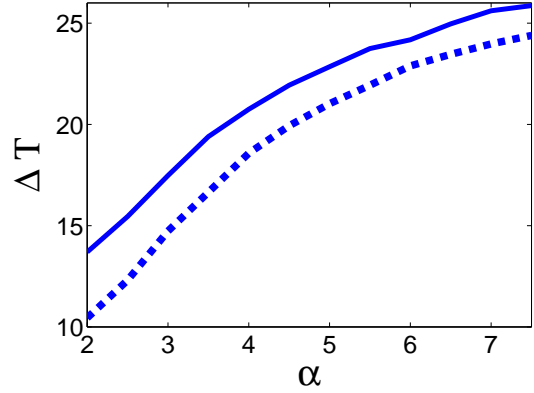}
   \caption{\label{fig1} Mean-value of the time delays between the two successive ionizations in Hamiltonian~(\ref{ham1}) for $I=4~{\rm PW}\cdot{\rm cm}^{-2}$. The other parameters are set as the ones in Ref.~\cite{pfei11} for a pulse duration of 33~fs. The continuous curve is for 3D calculations and the dashed curve corresponds to 2D calculations. }
\end{figure}
In Fig.~\ref{fig1}, we represent the time difference between the ionization of the two electrons for the parameters considered in Ref.~\cite{pfei11} (laser wavelength 788~nm, ellipticity 0.77, pulse duration 33~fs) and for an intensity of 4~${\rm PW}\cdot{\rm cm}^{-2}$. We see that the time difference is in agreement with the measured one (on the order of 10~fs) for $\alpha=2$, as in Ref.~\cite{zhou12}. However it increases as $\alpha$ is made larger, up to reaching a value comparable to the one obtained using soft-Coulomb potentials (on the order of 30~fs), both for the 2D and the 3D calculations. 
Therefore we argue that the quantitative agreement on ionization times is only coincidental. The relevance of adding the Heisenberg core potential has not been demonstrated. Note that a quantitative agreement with the experimental results has been obtained in Ref.~\cite{eber12} using soft-Coulomb potential, i.e., without invoking the Heisenberg core potential. We reached similar conclusions by using soft Coulomb potentials with varying effective charge for the electron-nucleus interaction and by decreasing the influence of the electron-electron correlation.   

In order to obtain qualitative agreement with experimental results (e.g., the knee shape of the double ionization probability versus intensity or the patterns of the ion momentum distribution), it has been shown in the past few decades that the specific choice of the potential does not matter. This means that the theoretical or numerical interpretation of these experiments is structurally stable. However using these models to obtain quantitative agreement is more subtle since it requires a fine tuning of the Hamiltonian model of the atom. This problem is central whether one works in the quantum or classical framework. 

C. Chandre$^1$, A. Kamor$^{1,2}$, F. Mauger$^1$, and T. Uzer$^2$\\
$^1$ Centre de Physique Th\'eorique, CNRS -- Aix-Marseille Universit\'e, Campus de Luminy, Case 907, 13288 Marseille cedex 09, France\\
$^2$ School of Physics, Georgia Institute of Technology, Atlanta, Georgia 30332-0430, USA

\end{document}